\documentstyle[manuscript,eqsecnum,aps,epsfig]{revtex}
\begin{document}


\preprint{}

\title{Double-Scaling Limit of a Broken Symmetry Quantum Field Theory}

\author{Carl M. Bender\cite{bye1}}
\address{Department of Physics, Washington University, St. Louis, MO 63130, USA}

\author{Stefan Boettcher\cite{bye2}}
\address{Department of Physics, Emory University, Atlanta, GA 30322, USA}

\author{H. F. Jones\cite{bye3}}
\address{Blackett Laboratory, Imperial College, London SW7 2BZ, UK}

\author{Peter N. Meisinger\cite{bye4}}
\address{Department of Physics, Washington University, St. Louis, MO 63130, USA}

\date{\today}

\maketitle

\begin{abstract}
The Ising limit of a conventional Hermitian parity-symmetric scalar quantum
field theory is a correlated limit in which two bare Lagrangian parameters, the
coupling constant $g$ and the {\it negative} mass squared $-m^2$, both approach
infinity with the ratio $-m^2/g=\alpha>0$ held fixed. In this limit the
renormalized mass of the asymptotic theory is finite. Moreover, the limiting
theory exhibits universal properties. For a non-Hermitian $\cal PT$-symmetric
Lagrangian lacking parity symmetry, whose interaction term has the form
$-g(i\phi)^N/N$, the renormalized mass diverges
in this correlated limit. Nevertheless, the asymptotic theory still has
interesting properties. For example, the one-point Green's
function approaches the value $-i\alpha^{1/(N-2)}$ independently of the
space-time dimension $D$ for $D<2$. Moreover, while the Ising
limit of a parity-symmetric quantum field theory is dominated by a dilute
instanton gas, the corresponding correlated limit of a $\cal PT$-symmetric
quantum field theory without parity symmetry is dominated by a constant-field
configuration with corrections determined by a weak-coupling expansion
in which the expansion parameter (the amplitude of the vertices of the
graphs in this expansion) is proportional to an inverse power of $g$. We thus
observe a
weak-coupling/strong-coupling duality in that while the Ising limit is a
strong-coupling limit of the quantum field theory, the expansion about this
limit takes the form of a conventional weak-coupling expansion. A
possible generalization of the Ising limit to dimensions $D<4$ is briefly
discussed.

\end{abstract}


\section{$\cal PT$-SYMMETRIC QUANTUM FIELD THEORY}
\label{s1}

Conventional field-theoretic Hamiltonians possess two crucial symmetries, the
continuous symmetry of the proper Lorentz group and the discrete symmetry of
Hermiticity. While Lorentz invariance is a physical requirement, Hermiticity is
a useful but rather mathematical constraint. However, assuming Lorentz
invariance and positivity of the spectrum of the Hamiltonian one can prove the
${\cal PCT}$ theorem and thereby establish the physical discrete symmetry of ${\cal PCT}$
invariance. Recent papers have investigated the consequences of imposing only
the physical symmetries of Lorentz invariance and ${\cal PCT}$ invariance in
constructing a Hamiltonian. The constraint of ${\cal PCT}$ invariance is weaker
than Hermiticity, so Hamiltonians having this property need not be Hermitian. In
quantum mechanics and in scalar quantum field theory the ${\cal C}$ operator is
unity, so ${\cal PCT}$ symmetry reduces to $\cal PT$ symmetry. While it has not
yet been proved, there is strong analytical and numerical evidence supporting
the conjecture that, except when ${\cal PT}$ symmetry is spontaneously broken,
the energy levels of many such Hamiltonians are all real and positive. The reality
and positivity of the spectrum is apparently a consequence of the ${\cal PT}$
symmetry of $H$. Hamiltonians having ${\cal PT}$ symmetry been studied in
quantum mechanics \cite{R1,R2,R9,R10,R11,R12,S1,S2,S3,S4,R13,S5,S6,S7} and in
quantum field theory \cite{R3,R4,R5,R6,R7,R8}.

A simple example of such a quantum-mechanical Hamiltonian is $H=p^2+ix^3$.
Hamiltonians of this form may be regarded as {\it complex deformations} of
conventional Hermitian Hamiltonians. To illustrate this deformation we consider
the Hamiltonian $H=p^2-(ix)^N$, where $N\geq2$ is a real number that is {\it not
necessarily an integer}. When $N=2$, we have the harmonic oscillator
Hamiltonian, whose spectrum is real and positive. As $N$ increases from $2$, the
entire spectrum of the Hamiltonian smoothly deforms as a function of $N$ and
remains real and positive for all values of $N>2$. Thus, these theories are in
effect the analytic continuation of conventional quantum mechanics into the
complex plane.

These non-Hermitian theories exhibit some remarkable properties. Most
interesting is that the expectation value of the operator $x$ in quantum
mechanics and the field $\phi$ in the corresponding quantum field theory is {\it
nonzero} when $N>2$. This is true even for the $p^2-x^4$ Hamiltonian that one
obtains at $N=4$, and it is also true for the $-g\phi^4$ scalar quantum field
theory. The $-g\phi^4$ quantum field theory is particularly surprising because
it has a positive real spectrum and exhibits a nonzero value of $\langle\phi
\rangle$, and in four-dimensional space-time has a dimensionless coupling
constant, is renormalizable, and is asymptotically free (and thus nontrivial).
It may thus provide a useful setting to describe the Higgs particle\cite{R7}.

In this paper we investigate the Euclidean scalar quantum field theory defined
by the Lagrangian density
\begin{equation}
{\cal L}={1\over2}(\nabla\phi)^2+{1\over2}m^2\phi^2-{g\over N}(i\phi)^N\quad(N>2
).
\label{e1.1}
\end{equation}
Our purpose here is to study this quantum field theory in the
correlated limit in which two bare Lagrangian parameters, the coupling
constant $g$ and the {\it negative} mass squared $-m^2$, both approach infinity
with the ratio
\begin{equation}
-m^2/g\equiv\alpha>0
\label{e1.2}
\end{equation}
held fixed. In a conventional parity-symmetric scalar quantum field theory this
limit is called the {\it Ising limit}. In the Ising limit the renormalized mass
of the asymptotic theory is finite. Moreover, the limiting theory exhibits
universal properties that will be discussed in Sec.~\ref{s2}. For
the non-Hermitian $\cal PT$-symmetric Lagrangian Eq.~(\ref{e1.1}) the
renormalized mass diverges in this correlated limit. We will show,
however, that the asymptotic theory exhibits intriguing properties. Of
considerable interest is that the one-point Green's function $G_1$ approaches
the finite value $-i\alpha^{1/(N-2)}$. Furthermore, while the Ising limit of a
parity-symmetric quantum field theory is dominated by a dilute instanton gas,
the corresponding correlated limit of a $\cal PT$-symmetric quantum field theory
lacking parity symmetry is dominated by a constant-field configuration with
corrections determined by a weak-coupling expansion in which the lines
represent propagators of the conventional weak-coupling form and the
vertices are proportional to an inverse power of $g$.

This paper is organized as follows. In Sec.~\ref{s2} we review the Ising limit
of a Hermitian parity-invariant self-interacting scalar quantum field theory and
consider this same correlated limit in a ${\cal
PT}$-symmetric quantum field theory. In Sec.~\ref{s3} we examine Hermitian and
non-Hermitian ${\cal PT}$-symmetric quantum field theories in the correlated
limit of Eq.~(\ref{e1.2}) for the special case of $D=0$. In Sec.~\ref{s4} we
investigate the one-dimensional case of Eq.~(\ref{e1.1}) in this correlated
limit by using the correspondence between one-dimensional field theory and
quantum mechanics. Finally, in Sec.~\ref{s5} we study this correlated limit for
a $D$-dimensional quantum field theory where $D<2$ and make some observations
concerning the case $D\geq2$.

\section{CONVENTIONAL ISING LIMIT OF SCALAR QUANTUM FIELD THEORY}
\label{s2}

The {\sl Ising limit} of a scalar quantum field theory is defined as follows.
Given the Lagrangian density for a $D$-dimensional Euclidean space quantum field
theory,
\begin{eqnarray}
{\cal L}={1\over 2}(\nabla\phi)^2+{1\over 2}m^2\phi^2+{g\over N}|\phi|^N
\quad(N>2),
\label{e2.1}
\end{eqnarray}
take the limit as the bare coupling constant $g\to\infty$ but demand that the
renormalized mass $M$ (the pole of the two-point Green's function) remain fixed
and finite. To satisfy this constraint the value of the bare mass squared $m^2$
must approach $-\infty$ so that the ratio $-m^2/g=\alpha$ is fixed. Thus, the
Ising limit is a correlated limit. In this limit the renormalized Green's
functions of Eq.~(\ref{e2.1}) approach universal $N$-independent values
\cite{ISING}.

The terminology ``Ising limit'' is taken from statistical mechanics. The Ising
model of statistical mechanics describes systems in which there are two equally
likely spin states. By analogy, in the correlated limits $g\to\infty$ and $m^2
\to-\infty$ the potential ${1\over 2}m^2\phi^2+{g\over N}|\phi|^N$ develops a
deep symmetric double well. In one Euclidean spacetime dimension (quantum
mechanics) the Lagrangian density ${\cal L}$ represents a particle that is
equally likely to be in one of two possible states, the left well or the right
well.

The Ising limit is a strong-coupling phenomenon and is not accessible by a
conventional perturbative treatment. However, a nonperturbative semiclassical
analysis in quantum mechanics can be used to calculate the amplitude for the
particle to tunnel from one well to the other. This tunneling amplitude is
exponentially small. The well is symmetric, so the splitting between the lowest
energy state and the first excited state is also exponentially small and is
proportional to this tunneling amplitude. The renormalized mass $M$ is the
difference between the energy of the (odd-parity) first excited state and the
energy of the (even-parity) ground state. The Ising limit exists because $M$ can
remain fixed even though the double-well potential becomes infinitely deep and
all of its energy levels approach negative infinity. The symmetry of the double
well is crucial; if the double well were not symmetric, the renormalized mass
could not remain finite as $g$ and $-m^2$ become large.

To determine the dimensionless renormalized Green's functions of a
$D$-dimensional Euclidean quantum field theory in the Ising limit, we follow a
routine procedure. First, we construct the vacuum persistence amplitude in the
presence of an external source:
\begin{eqnarray}
{\cal Z}[J]=\int{\cal D}\phi\exp\left\{-\int d^Dx\,[{\cal L}-J(x)\phi(x)]\right
\}.
\label{e2.3}
\end{eqnarray}
The connected unrenormalized $n$-point Green's functions are then obtained by
repeated functional differentiation with respect to the source $J$:
\begin{eqnarray}
G_n(x_1,x_2,x_3,\dots,x_n)=\left.{\delta\over\delta J(x_1)}{\delta\over\delta
J(x_2)}{\delta\over\delta J(x_3)}\cdots{\delta\over\delta J(x_n)}\ln({\cal Z}[J]
)\right|_{J=0}.
\label{e2.4}
\end{eqnarray}
(If the Lagrangian is symmetric under $\phi\to-\phi$, then Green's functions
having an odd number $n$ of legs vanish.) Alternatively, we may calculate these
Green's functions from the $n$-point correlation function,
\begin{eqnarray}
W_n(x_1,x_2,\ldots,x_n)\equiv{1\over{\cal Z}}\int{\cal D}\phi\,\phi(x_1)
\phi(x_2)\dots\phi(x_n)\exp\left(-\int dx\,{\cal L}\right),
\label{e2.5}
\end{eqnarray}
by subtracting the disconnected parts according to the formulas for the
cumulants:
\begin{eqnarray}
G_1(x_1) &=& W_1(x_1),\nonumber\\
G_2(x_1,x_2) &=& W_2(x_1,x_2)-W_1(x_1)W_1(x_2),\nonumber\\
G_3(x_1,x_2,x_3) &=& W_3(x_1,x_2,x_3)-W_1(x_1)W_2(x_2,x_3)-W_1(x_2)W_2(x_1,x_3)
\nonumber\\
&& \qquad -W_1(x_3)W_2(x_1,x_2)+2W_1(x_1)W_1(x_2)W_1(x_3),
\label{e2.6}
\end{eqnarray}
and so on.

Second, we construct the one-particle-irreducible (1PI)
connected Green's functions $\Gamma_n$ using the formulas in the Appendix,
specifically (\ref{ex.kk}), (\ref{ex.mm}), and (\ref{ex.oo}).
One effect of these relations is to
amputate the legs of the $n$-point unrenormalized Green's function by
multiplying by $(G_2)^{-n}$.

Third, we construct the dimensionless renormalized 1PI Green's functions
$\tilde\Gamma_n^{\rm ren}$. To do so we
perform a wave-function renormalization
by multiplying by $(\sqrt{Z})^n$, where $Z$ is the wave-function renormalization
constant. The {\sl dimensionless} renormalized Green's functions are then
obtained by multiplying by the appropriate power of the renormalized mass.
Normally $Z$ is
defined as the residue of the pole of the two-point Green's function.
However, in this paper we use the simpler
{\it intermediate renormalization scheme} in which the renormalization
is performed in momentum space with the Green's functions evaluated at zero
momentum on the external legs.
In this scheme the value of $Z$ is just the two-point Green's function in
momentum space multiplied by the square of the renormalized mass.

The dimensionless renormalized 1PI Green's functions
$\tilde\Gamma_n^{\rm ren}$ are the coefficients in the Taylor expansion of the
renormalized effective action. In the Ising limit of the parity-symmetric
Lagrangian density in Eq.~(\ref{e2.1}) these coefficients are known analytically
when the dimension $D$ of Euclidean spacetime is zero or one. The dimensionless
renormalized $2n$-point momentum-space Green's functions ${\tilde
\Gamma}_{2n}^{\rm ren}$ at zero external momentum are \cite{ISING}
\begin{eqnarray}
{\tilde \Gamma}_{2n}^{\rm ren}\Bigm|_{D=0} &=& -{n!\over 2n(2n-1)},\nonumber\\
{\tilde \Gamma}_{2n}^{\rm ren}(0,0,\ldots,0)\Bigm|_{D=1} &=& -{2^n\Gamma\left(
n-{1\over 2}\right)\over 4\Gamma\left({1\over 2}\right)}.
\label{e2.7}
\end{eqnarray}
Note that these results are independent of $N$; thus, apart from dimensional
dependence, the Ising limit is evidently universal.

In this paper we examine the Ising limit for the class of scalar quantum field
theories defined in Eq.~(\ref{e1.1}). While these theories are similar to those
in Eq.~(\ref{e2.1}), they do not possess parity symmetry. For such theories we
will show that in the correlated limit $g\to\infty$, $-m^2\to\infty$ with the
ratio in Eq.~(\ref{e1.2}) fixed, the Green's functions exhibit a remarkably
simple structure even though in this limit the renormalized mass now diverges.
Of course, when $N$ is not an even integer, the spectrum of a quantum field
theory whose interaction term is $\phi^N$ is not bounded below. Moreover, the
functional-integral representation for the vacuum persistence amplitude
\begin{eqnarray}
{\cal Z}=\int{\cal D}\phi\exp\left(-\int dx\,{\cal L}\right)
\label{e2.2}
\end{eqnarray}
does not exist. However, for the strange looking non-Hermitian Lagrangian
density in Eq.~(\ref{e1.1}), which was discussed in detail in \cite{R3}, it
appears that for $N\geq2$, the energy levels are all real and positive and that
for this Lagrangian density the functional integral in Eq.~(\ref{e2.2}) exists.
[Note that for $N\geq XXX$ the functional integral must be performed along a
complex contour that begins {\it below} the negative-real axis and ends {\it
below} the positive-real axis in the complex-$\phi$ plane. More precisely, the
contour approaches infinity within asymptotic wedges whose opening angles are
determined by the criterion that the functional integral in Eq.~(\ref{e2.2})
exist. These wedges are described in detail in Ref.~\cite{R3}.] The
$\cal PT$-symmetric Lagrangian density (\ref{e1.1}) is interesting because it
provides a simple model of a quantum field theory with a broken symmetry. When
$N=2$ the Lagrangian density represents a free theory, but as $N$ increases from
this value, the theory exhibits remarkable properties. For example, by direct
calculation one can show that the value of $\langle\phi\rangle$ is nonzero (it
has a negative-imaginary value) even if $N$ is an even integer\cite{R3}.

\section{CORRELATED LIMIT FOR $0$-DIMENSIONAL QUANTUM FIELD THEORY}
\label{s3}

In this section we discuss the Ising limit for zero-dimensional Euclidean
quantum field theory. The vacuum persistence amplitude ${\cal Z}$ and
correlation functions $W_n$ for such theories are expressible in terms of
conventional Riemann integrals.

\subsection{The parity-symmetric case}
\label{ss3a}

For the parity-symmetric theory the $n$-point correlation functions are
\begin{eqnarray}
W_n={\int_{-\infty}^{\infty}dt\,t^n\exp\left(-{1\over 2}m^2t^2-{g\over N}t^N
\right)\over\int_{-\infty}^{\infty}dt\,\exp\left(-{1\over 2}m^2t^2-{g\over N}t^N
\right)},
\label{e3.1}
\end{eqnarray}
where $N$ is an even integer greater than 2. Note that $W_n$ vanishes when $n$
is odd.

To evaluate the integrals in Eq.~(\ref{e3.1}) in the limit of large $g$ and
$-m^2$, we substitute $-m^2=\alpha g$, where $\alpha$ is a positive constant,
and use Laplace's method \cite{BO}. The Laplace points are the roots of
${d\over dt}\left({1\over 2}\alpha t^2-{1\over N}t^N\right)=\alpha t-t^{N-1}=0$.
Clearly, one Laplace point is always $t=0$ and expanding about this point gives
the usual weak-coupling Feynman perturbation series. However, as $g\to\infty$,
the contribution from this point vanishes exponentially relative to
contributions from other Laplace points. Two real Laplace points located at
$t=\pm \alpha^{1/(N-2)}$ dominate the asymptotic behavior of the integral
representation for $W_n$. We thus obtain the leading asymptotic behavior
\begin{eqnarray}
W_{2n}\sim \alpha^{2n/(N-2)} \qquad (g\to\infty).
\label{e3.3}
\end{eqnarray}

We then construct the connected Green's functions from the correlation functions
$W_{n}$ by using the zero-dimensional version of the cumulants in
Eq.~(\ref{e2.6}):
\begin{eqnarray}
G_1 &=& W_1,\nonumber\\
G_2 &=& W_2-(W_1)^2,\nonumber\\
G_3 &=& W_3-3W_1W_2+2(W_1)^3,\nonumber\\
G_4 &=& W_4-4W_1W_3-3(W_2)^2+12(W_1)^2W_2-6(W_1)^4,\nonumber\\
G_5 &=& W_5-5W_1W_4-10W_2W_3+20(W_1)^2W_3+30W_1(W_2)^2-60(W_1)^3W_2+24(W_1)^5,
\nonumber\\
G_6 &=& W_6-6W_1W_5-15W_2W_4+30(W_1)^2W_4-10(W_3)^2+120W_1W_2W_3-120(W_1)^3W_3
\nonumber\\
&&\qquad +30(W_2)^3-270(W_1)^2(W_2)^2+360(W_1)^4W_2-120(W_1)^6,
\label{e3.4}
\end{eqnarray}
and so on. In the symmetric case these equations simplify enormously because
$W_{2n+1}=0$.

We recover the result in Eq.~(\ref{e2.7}) by substituting Eq.~(\ref{e3.3}) into
Eq.~(\ref{e3.4}) and then following the renormalization procedure described in
Sec.~\ref{s2}: We amputate the external legs, and then multiply by the
appropriate power of the renormalized mass $M$, where $G_2=M^{-2}$, to make
the Green's functions dimensionless. (It is not necessary to perform a
wave-function renormalization in zero dimensions; one can always choose the
wave-function renormalization constant $Z$ to be unity.) Note that the
dimensionless renormalized Green's functions in (\ref{e2.7}) are
pure numbers that do not depend on the value of the constant $\alpha$.

\subsection{The parity-nonsymmetric case}
\label{ss3b}

Now consider the zero-dimensional version of the scalar quantum field theory in
Eq.~(\ref{e1.1}). For this theory the $n$-point correlation functions $W_n$ are
\begin{eqnarray}
W_n={\int_{-\infty}^{\infty}dt\,t^n\exp\left[-{1\over 2}m^2t^2+{g\over N}(it)^N
\right]\over\int_{-\infty}^{\infty}dt\,\exp\left[-{1\over 2}m^2t^2+{g\over N}
(it)^N\right]}.
\label{e3.5}
\end{eqnarray}

To evaluate $W_n$ we split the range of integration in each of these integrals
as a sum of two contributions:
\begin{eqnarray}
\int_{-\infty}^\infty dt\,t^n\cdots=\int_{-\infty}^0 dt\,t^n\cdots+\int_0^\infty
dt\,t^n\cdots=2\left({{\rm Re}~({\rm if}~n~{\rm even})\atop i\,{\rm Im}~({\rm if
}~n~{\rm odd})}\right)\int_0^\infty dt\,t^n\cdots.
\label{e3.6}
\end{eqnarray}
This integral exists if $1<{\rm Re}\,N<3$.

Note that $W_n$ is an analytic function of $N$ for $N\geq0$ because the
region inside of which the
integration path in Eq.~(\ref{e3.5}) lies
{\em is an implicit function of} $N$.
Indeed, as $N$ ranges through real values, the paths of integration of the two
integrals in Eq.~(\ref{e3.6}) lie inside wedge-shaped regions
that {\em rotate in opposite directions} \cite{ROT}.
It is convenient to take the paths of integration to lie at the center of
the wedges. In this case, the
path of integration of the first integral connects complex $\infty$ to 0 in
the $t$ plane along the straight line
\begin{eqnarray}
{\rm path}~1:\quad{\rm arg}\,t=-\pi/2-\pi/N.
\label{e3.8}
\end{eqnarray}
The second integration path runs from $0$ to complex $\infty$ along
\begin{eqnarray}
{\rm path}~2:\quad {\rm arg}\,t=-\pi/2+\pi/N.
\label{e3.9}
\end{eqnarray}
The opening angle of each wedge is $\pi/N$.

When $N=2$ (free field theory),
the wedges
are centered about the positive and negative real axes and the opening
angle of the wedges is
$90^\circ$. In this case
path 1 connects $-\infty$ to $0$ and path $2$
connects $0$ to $\infty$ along the real-$t$ axis. Here,
$W_n$ is {\em
real} and parity symmetry is unbroken. As $N$ increases, path $1$ rotates
anticlockwise and path $2$ rotates clockwise.
Integration along the real axis is no longer allowed when $N\geq3$.
The two paths slope
downward at $45^\circ$ angles when $N=4$ ($\phi^4$ field theory).
For all $N>2$ we find that $W_{2n+1}\neq0$, demonstrating that parity
symmetry is broken.

Now we discuss the Ising limit of the theory. To analyze the asymptotic
behavior of the integrals in Eq.~(\ref{e3.5}) we examine the expression
\begin{equation}
L(t)={1\over 2}m^2t^2-{g\over N}(it)^N\quad(N>2).
\label{e3.11}
\end{equation}
The saddle points that determine
the asymptotic behavior are the zeros of $L'(t)=m^2t-ig(it)^{N-1}$.
Remember that both $g$ and $-m^2$ are large such that the ratio $\alpha=-m^2/g$ is
fixed. There are many roots of $L'(t)=0$: First there is a root at $t=0$, which
is the perturbative root, corresponding to expansions in powers of $g$. Second,
there is a ring of roots surrounding the origin. The most important of these
roots, and the one that determines the Ising limit of the theory, lies on the
negative imaginary axis:
\begin{eqnarray}
t_0=-i\alpha^{1/(N-2)}.
\label{e3.17}
\end{eqnarray}

\begin{figure}[t]
\vspace{5.0in}
\includegraphics{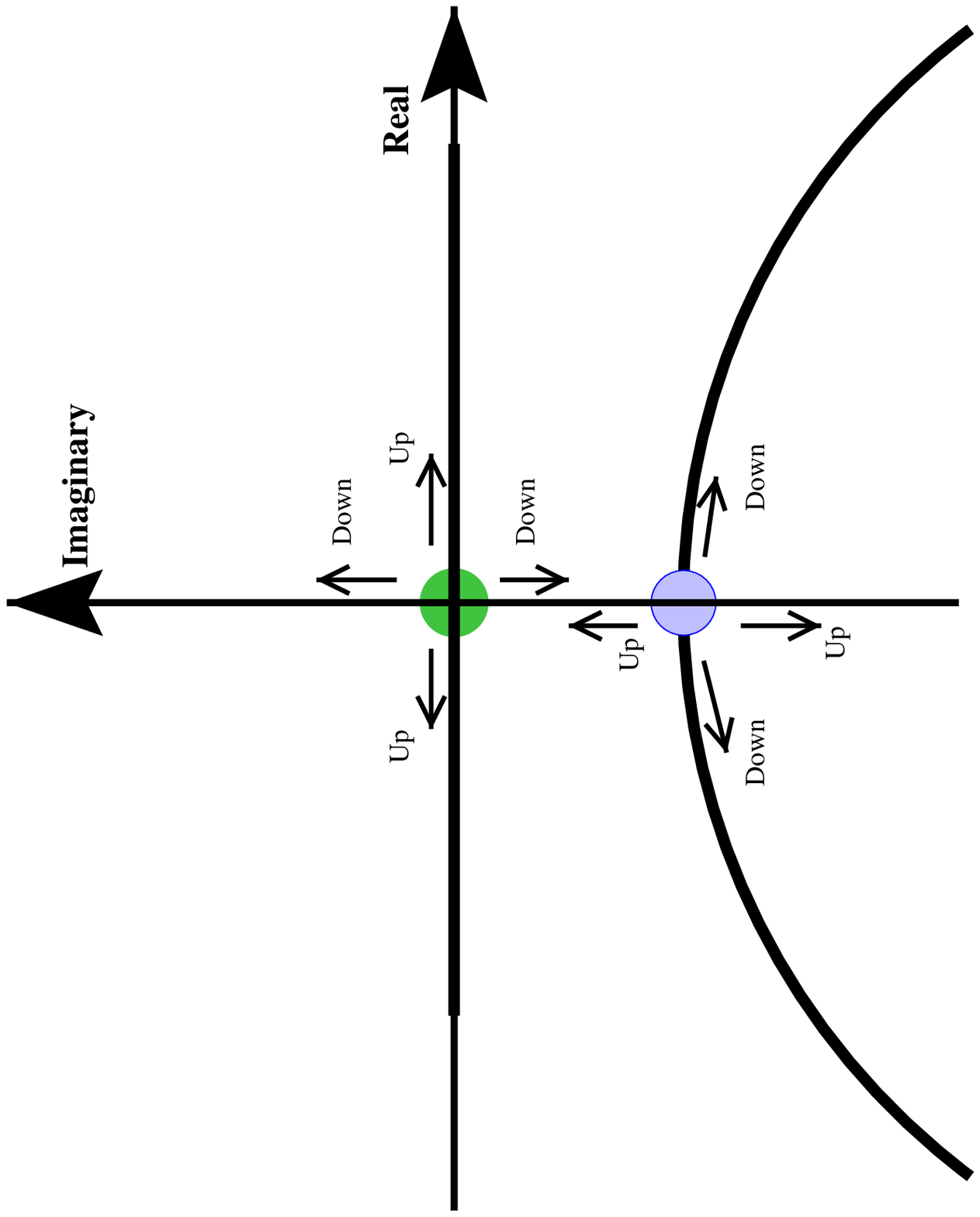}
\caption{Directions
of the saddle points used for the zero-dimensional Ising limit of
a PT-symmetric theory.} \label{f1}
\end{figure}

To find the directions of the saddle points we calculate the second derivative
of $L$: $L''(t)=m^2+g(N-1)(it)^{N-2}$. Thus, $L''(0)=m^2$ and $L''(t_0)=-(N-2)
m^2$, which is positive because $m^2$ is negative and $N>2$. Hence, the down
directions for the saddle point at $t=0$ go vertically along the imaginary axis.
The down directions for the saddle point $t_0$ are locally horizontal (see
Fig.~\ref{f1}). As we trace the down paths away from the saddle point $t_0$ they
curve downward and align with the directions in Eqs.~(\ref{e3.8}) and
(\ref{e3.9}). This verifies that $t_0$ is the saddle point that we should use.

It is straightforward to find the leading asymptotic behavior of the integrals
in Eq.~(\ref{e3.5}). The Gaussian corrections cancel and we obtain the
leading-order result
\begin{eqnarray}
W_n \sim (t_0)^n.
\label{e3.19}
\end{eqnarray}
However, when we substitute this result into the formulas in Eq.~(\ref{e3.4}),
we find that except for $G_1$, each of the Green's functions {\em vanishes} to
leading order. This happens because the sum of the numerical coefficients in
each cumulant except the first is zero. (This does not happen in the
parity-symmetric case because $W_n\equiv 0$ for odd $n$.)

Therefore, we are forced to perform the asymptotic analysis to higher order. For
example, to obtain the first nonvanishing contribution to $G_2$ we must
calculate $W_1$ and $W_2$ to one order beyond the Gaussian approximation; to
obtain $G_3$ we must calculate $W_1$, $W_2$, and $W_3$ to two orders beyond the
Gaussian approximation; to obtain $G_4$ we must calculate $W_1$, $W_2$, $W_3$,
and $W_4$ to three orders beyond the Gaussian approximation; and so on. To
perform this calculation we will require the $k$th derivative of $L(t)$ in
(\ref{e3.11}):
\begin{eqnarray}
L^{(k)}(t)=-{i^kg\Gamma(N)\over\Gamma(N-k+1)}(it)^{N-k}\qquad (k\geq 3).
\label{e3.20}
\end{eqnarray}
Substituting the saddle point $t=t_0$ gives
\begin{eqnarray}
L^{(k)}(t_0)=-{i^kg\Gamma(N)\over\Gamma(N-k+1)}\alpha^{N-k\over N-2}
\qquad(k\geq3).
\label{e3.21}
\end{eqnarray}

The expression for $W_n$ then has the form
\begin{eqnarray}
W_n \sim { \int dt\, t^n\,
\exp\left[\sum_{k=2}^{\infty}-{1\over k!}L^{(k)}(t_0)(t-t_0)^k\right]
\over
\int dt\,\exp\left[\sum_{k=2}^{\infty}-{1\over k!}L^{(k)}(t_0)(t-t_0)^k\right]}.
\label{e3.22}
\end{eqnarray}
Note that $L(t_0)$ cancels from the numerator and denominator. Next, we make the
translation $s=t-t_0$ and the scaling $s=v\epsilon\alpha^{1/(N-2)}$, where
$\epsilon^2=\alpha^{-N/(N-2)}/[g(N-2)]$. The result is
\begin{eqnarray}
W_n\sim (t_0)^n{\int dv\, (1+iv\epsilon)^n\, e^{-{1\over 2}v^2}\exp\left[
\sum_{k=3}^{\infty}{v^k i^k\Gamma(N)\epsilon^{k-2}\over k!\Gamma(N+1-k) (N-2)}
\right] \over\int dv\,e^{-{1\over 2}v^2}\exp\left[\sum_{k=3}^{\infty}{v^k i^k
\Gamma(N)\epsilon^{k-2}\over k!\Gamma(N+1-k) (N-2)}\right]}.
\label{e3.23}
\end{eqnarray}
We expand the integrands in the numerator and denominator as series in powers of
$\epsilon$ and perform the Gaussian integrals. We then substitute the result
into Eq.~(\ref{e3.4}) to obtain the small-$\epsilon$ leading asymptotic
approximations to the unrenormalized connected Green's functions $G_n$:
\begin{eqnarray}
G_1 &\sim& t_0,\nonumber\\
G_2 &\sim& -(t_0)^2\epsilon^2,\nonumber\\
G_3 &\sim& -(t_0)^3\epsilon^4 (N-1),\nonumber\\
G_4 &\sim& -2(t_0)^4\epsilon^6 (N-1)N,\nonumber\\
G_5 &\sim& -3(t_0)^5\epsilon^8 (N-1)(N+1)(2N-1),\nonumber\\
G_6 &\sim& -8(t_0)^6\epsilon^{10}(N-1)N(N+2)(3N-2),
\label{e3.24}
\end{eqnarray}
and so on. The general formula for $n>2$ is
\begin{eqnarray}
G_n\sim -(t_0)^n\epsilon^{2n-2}(N-1)(N-2)^{n-2}\Gamma\left[{(n-2)(N-1)\over
N-2}\right]\Big/\Gamma\left({n-2\over N-2}\right).
\label{e3.25}
\end{eqnarray}

Next we construct the coefficients of the effective action. Using the
relations given in the Appendix we obtain
\begin{eqnarray}
\Gamma_1 &\sim&-(t_0)^{-1}\epsilon^{-2},\nonumber\\
\Gamma_2 &\sim&-(t_0)^{-2}\epsilon^{-2}(N-2),\nonumber\\
\Gamma_3 &\sim&-(t_0)^{-3}\epsilon^{-2}(N-1)(N-4),\nonumber\\
\Gamma_4 &\sim&-(t_0)^{-4}\epsilon^{-2}(N-1)(N-3)(N-5),\nonumber\\
\Gamma_5 &\sim&-(t_0)^{-5}\epsilon^{-2}(N-1)(N-3)(N-4)(N-6),\nonumber\\
\Gamma_6 &\sim&-(t_0)^{-6}\epsilon^{-2}(N-1)(N-3)(N-4)(N-5)(N-7),
\label{e3.26}
\end{eqnarray}
and so on. The general formula is
\begin{eqnarray}
\Gamma_n\sim-(t_0)^{-n}\epsilon^{-2}{1\over(N-2)}\,\left[
{(N-n-1)\Gamma(N)\over\Gamma(N-n+1)}+\delta_{n,2}\right].
\label{e3.27}
\end{eqnarray}

To construct the {\it
dimensionless} renormalized coefficients of the effective action $\tilde
\Gamma_n^{\rm ren}$ by multiplying $\Gamma_n$ by the appropriate power of the
renormalized mass $M^2=1/G_2=-(\epsilon t_0)^{-2}=\alpha g(N-2)$ according to
$\tilde\Gamma_n^{\rm ren}=\Gamma_n (M^2)^{-n/2}$.
Thus, we have the result
\begin{eqnarray}
\tilde\Gamma_n^{\rm ren}\sim-\epsilon^{n-2}{i^n\over(N-2)}\,\left[
{(N-n-1)\Gamma(N)\over\Gamma(N-n+1)}+\delta_{n,2}\right],
\label{e3.29}
\end{eqnarray}
independent of $t_0$.

\section{CORRELATED LIMIT FOR THE SCHR\"ODINGER EQUATION}
\label{s4}

The Lagrangian in Eq.~(\ref{e1.1}) is a field-theoretic generalization of the
quantum-mechanical theory described by the non-Hermitian Hamiltonian
\begin{eqnarray}
H={1\over 2}p^2+{1\over 2}x^2-{g\over N}(ix)^N.
\label{e4.1}
\end{eqnarray}
This Hamiltonian is ${\cal PT}$ symmetric because under parity reflection ${\cal
P}:~p\to-p$ and $x\to-x$ and under time reversal, which is an antiunitary
operation, ${\cal T}:~p\to-p$, $x\to x$, and $i\to-i$. The Ising substitution
$m^2=-\alpha g$ gives the Schr\"odinger equation
\begin{eqnarray}
-{1\over 2}\psi''(x)+\left[-{1\over 2}\alpha g x^2-{g\over N}(ix)^N
-E\right]\psi(x)=0,
\label{e4.2}
\end{eqnarray}
where $\alpha>0$. Note that the eigenvalue problem is posed on a path in the
complex-$x$ plane and that the endpoints of this path lie in complex wedges
similar to the wedges discussed earlier for the complex path integral. See
Refs.~\cite{R1} and \cite{R2}.

We seek the large-$g$ behavior of (\ref{e4.2}). In this limit the
energy $E$ scales like $g$ because the potential scales like $g$ (this will be
verified shortly). Thus, we make the substitution
\begin{eqnarray}
E=\lambda g.
\label{e4.3}
\end{eqnarray}
We now can study $Q(x)=-{1\over 2}\alpha x^2-{1\over N}(ix)^N -\lambda$ in the
resulting Schr\"odinger equation
\begin{eqnarray}
-{1\over 2}\psi''(x)+g Q(x)\psi(x)=0.
\label{e4.4}
\end{eqnarray}

For large $g$ it is the turning points [the zeros of $Q(x)$] that determine the
physics of the problem. More precisely, it is the lowest-lying pair of turning
points that control the physics. Near
this pair of turning points, the polynomial $Q(x)$ can be approximated by a
parabola. To construct the parabola, we locate the point on
the imaginary axis midway between the pair of turning points by differentiating
$Q(x)$ and setting $Q'(x)=-\alpha x-i(ix)^{N-1}=0$. The value of $x$ on the
negative imaginary axis that solves this equation is
$x_0=-i\alpha^{1/(N-2)}$.
At this value of $x$, we see that $Q(x)$ vanishes if
\begin{eqnarray}
\lambda= {N-2\over2N}\alpha^{N\over N-2},
\label{e4.8}
\end{eqnarray}
which justifies the scaling of $E$ used in (\ref{e4.3}).

Next, we expand the Schr\"odinger equation (\ref{e4.4}) around the point $x_0$
by substituting
\begin{eqnarray}
x=-i\alpha^{1\over N-2}+\epsilon t\quad{\rm and}\quad\lambda={N-2\over2N}
\alpha^{N\over N-2}+\delta.
\label{e4.9}
\end{eqnarray}
Here, we treat $\epsilon$ and $\delta$ as small parameters, whose size will
be determined below. We obtain
\begin{eqnarray}
-{1\over 2\epsilon^2}\psi''(t)+g \left[-{N-2\over2}\alpha \epsilon^2 t^2
+i{(N-1)(N-2)\over6}\alpha^{N-3\over N-2}\epsilon^3 t^3 -\delta\right]\psi(t)=0.
\label{e4.10}
\end{eqnarray}

The requirement of dominant balance \cite{BO} implies that we must choose
\begin{eqnarray}
\epsilon=g^{-1/4}\qquad{\rm and}\qquad\delta=\beta g^{-1/2},
\label{e4.11}
\end{eqnarray}
where $\beta={\rm O}(1)$ is a constant. To leading order the resulting
Schr\"odinger equation reads
\begin{eqnarray}
-{1\over 2}\psi''(t)+\left[-{N-2\over2}\alpha t^2
+i{(N-1)(N-2)\over6}\alpha^{N-3\over N-2}g^{-1/4} t^3 -\beta\right]\psi(t)=0.
\label{e4.12}
\end{eqnarray}
As $g\to\infty$ this equation becomes the eigenvalue problem for the harmonic
oscillator, whose $n$th eigenvalue is $\beta=(n+1/2)\sqrt{(N-2)\alpha}$, where
$n=0,1,2,\ldots$ is an integer. Thus, for the Schr\"odinger equation
(\ref{e4.2}) the $n$th eigenvalue in the Ising limit is
\begin{eqnarray}
E_n={N-2\over2N}\alpha^{N\over N-2}g+(n+1/2)\sqrt{(N-2)\alpha g}
\quad(n=0,1,2,\ldots)
\label{e4.13}
\end{eqnarray}
with higher-order corrections of order $g^0$.

From this formula we can determine the renormalized mass $M_{\rm R}$:
\begin{eqnarray}
M_{\rm R}\equiv E_1-E_0=\sqrt{(N-2)\alpha g}.
\label{e4.14}
\end{eqnarray}
Observe that unlike the conventional Ising  limit, the renormalized mass
{\it diverges} as $g\to\infty$. Thus, the unrenormalized two-point Green's
function, which behaves like $M_{\rm R}^{-2}$, vanishes as $g\to\infty$
like $1/g$, in agreement with the result in (\ref{e3.24}) for the
zero-dimensional field theory.

We can determine the one-point Green's function $G_1$ by calculating the
expectation value of $x$ in the ground-state wave function. Specifically, we
calculate the ratio
\begin{eqnarray}
G_1\equiv{\int dx\,x[\psi_0(x)]^2 \over \int dx\,[\psi_0(x)]^2},
\label{e4.15}
\end{eqnarray}
where we obtain the ground-state wave function $\psi_0$ by setting $g=\infty$ in
(\ref{e4.12}). Because $\psi_0$ is a Gaussian in
$t$ and $x=x_0+\epsilon t$ from (\ref{e4.9}), we immediately have
\begin{eqnarray}
G_1=x_0= -i\alpha^{1\over N-2},
\label{e4.16}
\end{eqnarray}
which has a finite negative-imaginary value as $g\to\infty$. This result is
identical to that obtained in (\ref{e3.24}) for the $D=0$ theory.

To calculate $G_1$ to higher order we need to solve
$-{1\over2}\psi''(s)+\left[-{1\over2}+{1\over2}s^2+i\eta s^3\right]\psi(s)=0$
as a perturbation series
$$\psi(s)=e^{-s^2/2}[1+\eta f(s)+{\rm O}(\eta^2)].$$
We find that $f(s)=-i(s+s^3/3)$. Finally, we use this result in the integral
(\ref{e4.15}) to obtain
\begin{eqnarray}
G_1=-i\alpha^{1\over N-2}\left(1+{N-1\over4\sqrt{N-2}}\alpha^{-{N+2\over2N-4}}
g^{-1/2}\right).
\label{e4.17}
\end{eqnarray}

\section{CORRELATED LIMIT FOR $D$-DIMENSIONAL QUANTUM FIELD THEORY}
\label{s5}

In this section we use functional-integral techniques to study the Ising limit
in a scalar self-interacting scalar Euclidean quantum field theory of space-time
dimension $D<2$. We focus on the calculation of the one-point Green's function
$G_1$:
\begin{eqnarray}
G_1\equiv{\int{\cal D}\phi\,\phi(0)\exp\left(-\int d^Dx\,{\cal L}\right)\over
\int{\cal D}\phi\,\exp\left(-\int d^Dx\,{\cal L}\right)},
\label{e5.1}
\end{eqnarray}
where the Lagrangian density $\cal L$ is given in (\ref{e1.1}). Making the
substitution in (\ref{e1.2}) and letting $x=s/\sqrt{g}$ gives the result
\begin{eqnarray}
G_1={\int{\cal D}\phi\,\phi(0)\exp\left(-g^{1-D/2}S[\phi]\right)\over
\int{\cal D}\phi\,\exp\left(-g^{1-D/2}S[\phi]\right)},
\label{e5.2}
\end{eqnarray}
where
\begin{eqnarray}
S[\phi]=\int d^Ds\,\left[{1\over2}(\nabla\phi)^2-{1\over2}\alpha\phi^2-{1\over
N}(i\phi)^N\right]\quad(N>2).
\label{e5.3}
\end{eqnarray}

If we assume that $D<2$, then as $g\to\infty$ we can use saddle-point
methods to determine the behavior of $G_1$ in (\ref{e5.2}) as $g\to\infty$
because the coefficient of $S[\phi]$ is large.\footnote{Because the
parameter $g^{1-D/2}$ carries dimensions we cannot really interpret this
parameter as being large. Thus, the saddle-point expansion of the integral
is a formal perturbative procedure. We will see later on [see Eq.~(\ref{e5.23})]
that the actual
dimensionless large parameter is $(\alpha g)^{1-D/2}\alpha^{2/(N-2)}$.
This parameter can be large even if $D>2$ if we redefine
the correlated Ising limit (\ref{e1.2}) by regarding $\alpha$ as a large
parameter rather than a fixed constant.} We begin by taking the
functional derivative of $S[\phi]$. The saddle points are determined by the
equation
\begin{eqnarray}
{\delta\over\delta\phi(t)}S[\phi]=-\nabla^2\phi(t)-\alpha\phi(t)-i\left[i\phi(t)
\right]^{N-1}=0.
\label{e5.4}
\end{eqnarray}
The solution to this equation is a saddle point at $\phi=0$ and a ring of
saddle points centered about $0$. The dominant saddle point is the one on
the negative imaginary axis:
\begin{eqnarray}
\phi_0=-i\alpha^{1\over N-2}.
\label{e5.5}
\end{eqnarray}
The complex contour can be connected to this saddle point.
If we substitute the value of $\phi_0$, we get
\begin{eqnarray}
G_1\sim -i\alpha^{1\over N-2} \quad(g\to\infty).
\label{e5.6}
\end{eqnarray}

We now calculate all higher-order corrections. To do so we substitute
\begin{eqnarray}
\phi(s)=\phi_0+\eta(s)=-i\alpha^{1\over N-2}+\eta(s),
\label{e5.7}
\end{eqnarray}
where we treat $\eta(s)$ as small; that is, $\eta(s)<<1$. To illustrate
the procedure, we expand the functional $S$ in (\ref{e5.3}) to third
order in powers of $\eta(s)$. The result is
\begin{eqnarray}
S_3[\eta]=\int d^Ds\,\left[{N-2\over2N}\alpha^{N\over N-2}+{1\over2}(\nabla\eta)
^2+{N-2\over2}\alpha\eta^2+{(N-1)(N-2)\over6}\alpha^{N-3\over N-2}i\eta^3
\right].
\label{e5.8}
\end{eqnarray}
We can now rewrite (\ref{e5.2}) in the form
\begin{eqnarray}
G_1= -i\alpha^{1\over N-2}+
{\int{\cal D}\eta\,\eta(0)\exp\left(-g^{1-D/2}S_3[\eta]\right)\over
\int{\cal D}\eta\,\exp\left(-g^{1-D/2}S_3[\eta]\right)},
\label{e5.9}
\end{eqnarray}
The constant term in $S_3$, which is proportional to the volume of
Euclidean space-time, cancels from the exponentials in the numerator and the
denominator in (\ref{e5.9}). Treating $\eta$ as small, we expand the cubic term
in the exponential as a power series in $\eta^3$ and keep the first nontrivial
term. In the denominator, the term proportional to $\eta^3$ vanishes by oddness,
but in the numerator we retain the cubic term because the leading term vanishes
by oddness:
\begin{eqnarray}
G_1&\sim&-i\alpha^{1\over N-2}-i{(N-1)(N-2)\over6}\alpha^{N-3\over N-2}g^{1-D/2}
\nonumber\\
&&\quad\times{\int{\cal D}\eta\,\eta(0)\int d^Dt\,\eta^3(t)\exp\left(-g^{1-D/2}
S_{\rm free}[\eta]\right)\over \int{\cal D}\eta\,\exp\left(-g^{1-D/2}
S_{\rm free}[\eta]\right)},
\label{e5.10}
\end{eqnarray}
where $S_{\rm free}[\eta]={1\over2}\int d^Ds\,\left[(\nabla\eta)^2+(N-2)\alpha
\eta^2\right]$.

We can evaluate this ratio of functional integrals {\it exactly}. To do so
we introduce an external source function $J(s)$ in the integral in the
numerator:
\begin{eqnarray}
G_1&\sim&-i\alpha^{1\over N-2}-i{(N-1)(N-2)\over6}\alpha^{N-3\over N-2}
g^{-3+3D/2} \nonumber\\
&&\quad\times{\delta\over\delta J(0)}
\int d^Dt\left[{\delta\over\delta J(t)}\right]^3
{\int{\cal D}\eta\, \exp\left(-g^{1-D/2}S_J[\eta]\right)\over
\int{\cal D}\eta\,\exp\left(-g^{1-D/2}S_{\rm free}[\eta] \right)}\Biggm|_{J=0},
\label{e5.12}
\end{eqnarray}
where $S_J[\eta]=\int d^Ds\,\left({1\over2}([\nabla\eta(s)]^2+{1\over2}\alpha
(N-2)[\eta(s)]^2 -J(s)\eta(s)\right)$.

Next, we evaluate the Gaussian integral in the numerator by completing the
square. Upon doing so, the integral in the denominator cancels and what
remains is the formula
\begin{eqnarray}
G_1&\sim& -i\alpha^{1\over N-2}-i {(N-1)(N-2)\over6}\alpha^{N-3\over N-2}
g^{-3+3D/2}\nonumber\\
&&\quad\times{\delta\over\delta J(0)}
\int d^Dt\left[{\delta\over\delta J(t)}\right]^3 \exp\left({1\over2}g^{1-D/2}
\int\int dr\,ds\, J(r)J(s)\Delta(r,s)\right)\Bigm|_{J=0},
\label{e5.14}
\end{eqnarray}
where $\Delta$ is the coordinate-space propagator satisfying the
Euclidean coordinate space Green's function equation
\begin{eqnarray}
\left[\nabla^2 +\alpha (N-2)\right]\Delta(r,s)=\delta(r-s).
\label{e5.15}
\end{eqnarray}

The final step is to expand the exponential containing the external source
and to perform the indicated differentiations. The result is
\begin{eqnarray}
G_1\sim -i\alpha^{1\over N-2}-i {(N-1)(N-2)\over2}\alpha^{N-3\over N-2}
g^{-1+D/2} \int dr\,\Delta(r,0)\Delta(0,0).
\label{e5.16}
\end{eqnarray}
This expression has a graphical interpretation: A propagator connects the origin
to the point $r$ where there is a tadpole. In
momentum space the propagator is
${\tilde\Delta}(p)=[{1\over p^2+\alpha(N-2)}]^{-1}$ and
\begin{eqnarray}
\Delta(r,s)=(2\pi)^{-D}\int d^Dp\,e^{ip(r-s)}{1\over p^2+\alpha(N-2)}.
\label{e5.18}
\end{eqnarray}
Thus, ${\tilde\Delta}(0)=\int dr\,\Delta(r,0)={1\over\alpha(N-2)}$ and $\Delta
(0,0)=\Gamma\left(1-{D\over2}\right)(4\pi)^{-D/2}[\alpha(N-2)]^{-1+D/2}$. Thus,
our final result for the one-point Green's function is
\begin{eqnarray}
G_1\sim -i\alpha^{1\over N-2}\left[1+{N-1\over2}[g(N-2)]^{-1+D/2}(4\pi)^{-D/2}
\Gamma\left(1-{D\over2}\right)\alpha^{{D\over2}-{N\over N-2}}\right],
\label{e5.19}
\end{eqnarray}
which agrees exactly with (\ref{e3.24}) for the case $D=0$ and (\ref{e4.17})
for the case $D=1$.

\begin{figure}[t]
\centerline{\epsfxsize=4.5in\epsffile{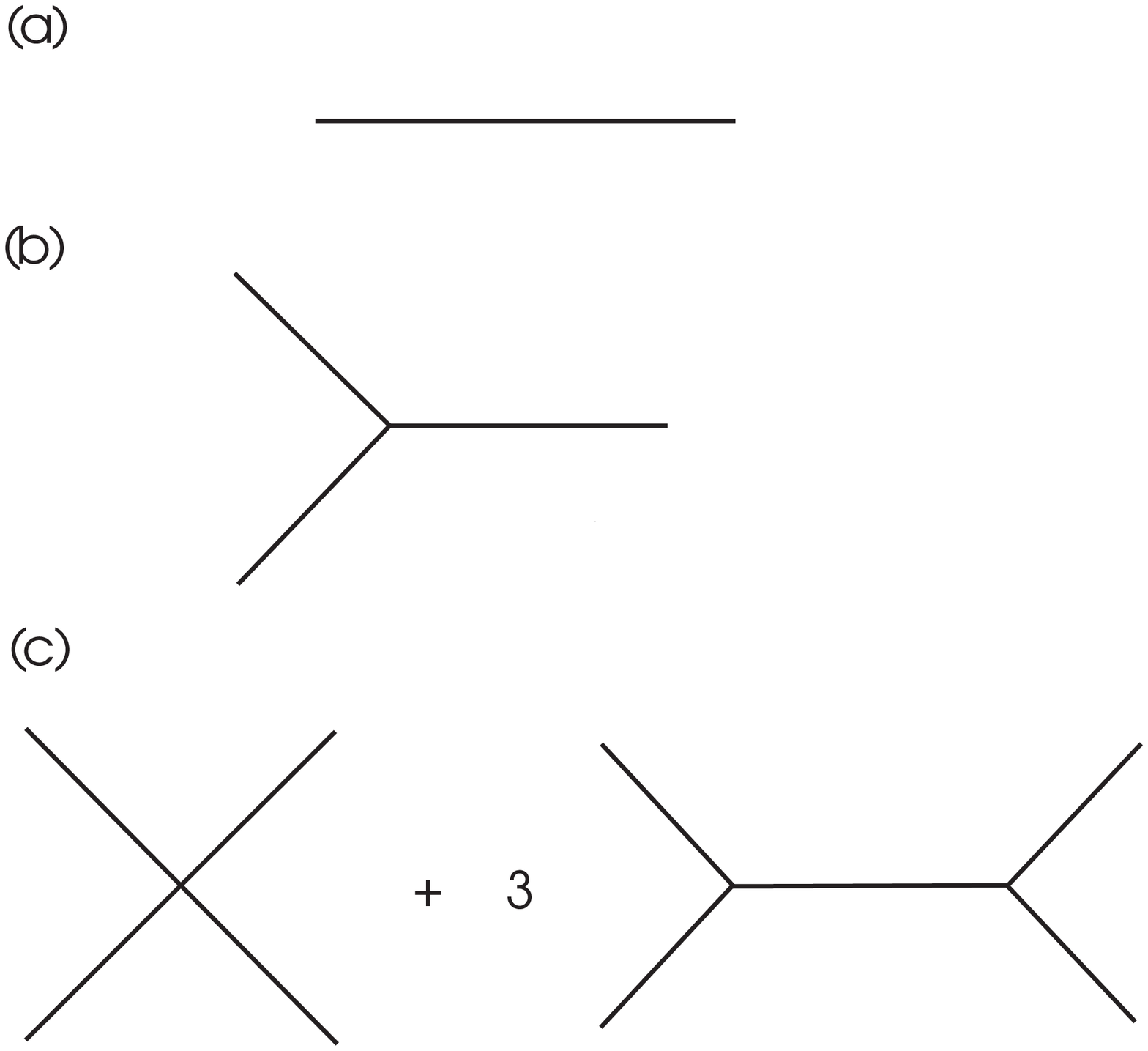}}
\vspace{1in} \caption{Tree graphs of the dual expansion
contributing to the asymptotic form of the Green's functions.
Graph (a) is the leading contribution to $G_2$, as given in
Eq.~(\protect\ref{e5.20}). Graph (b) is the leading contribution
to $G_3$, and graphs (c) are the two leading contributions to
$G_4$.} \label{f2}
\end{figure}

The procedure we have used to calculate $G_1$ can be generalized
to calculate any of the Green's functions $G_n$ by taking advantage of the
graphical methods developed above. For the two-point Green's function we
immediately obtain to leading order
\begin{eqnarray}
G_2(x,y)\sim g^{-1+D/2}\Delta(x\sqrt{g},y\sqrt{g}),
\label{e5.20}
\end{eqnarray}
which in momentum space gives
\begin{eqnarray}
G_2(p)\sim {1\over p^2+(N-2)\alpha g}.
\label{e5.21}
\end{eqnarray}
From this equation we see that to leading order the renormalization constant
$Z=1$ and that the renormalized mass $M_{\rm R}$ is given by
\begin{eqnarray}
M_{\rm R}^2=(N-2)\alpha g.
\label{e5.22}
\end{eqnarray}
Observe that this result is independent of the dimension $D$ and agrees with
the result for $D=0$ and also with that in (\ref{e4.14}) for $D=1$,
which was derived by solving the Schr\"odinger equation.
Figure \ref{f2} shows the tree graphs contributing to $G_2$
through $G_4$.

In fact, we find that the results in Eq.~(\ref{e3.24}) are valid
for any dimension $D<2$, provided that they are interpreted as
momentum space Green's functions evaluated at zero momentum on all
external legs. The parameter $\epsilon^2$ is, however, dimensional
for $D\neq 0$. Eqs.~(\ref{e3.25}) and (\ref{e3.27}) for the 1PI
Green's functions are also valid with the same understanding.

Thus, we observe a form of universality; the expressions for the
Green's functions are the same for all $D$ and only depend
on  $N$, the exponent in the interaction term. This
is quite different from the usual statement of universality, in
which the Green's functions are independent of $N$ but do depend
on the value of $D$. However, Eq.~(\ref{e3.29}) is $D$ dependent
insofar as the parameter $\epsilon$ must be replaced by its
dimensionless version, given by
\begin{eqnarray}
\tilde\epsilon^2= {\alpha^{{D\over 2}-{N\over N-2}}\over
[(N-2)g]^{1-{D\over 2}}}. \label{e5.23}
\end{eqnarray}
This is the natural small parameter governing the asymptotic
expansion, as in Eq.~(\ref{e5.19}).

In the calculations performed so far it has been assumed implicitly
that $D<2$. Indeed, $D$ cannot exceed $2$ if $\alpha$ is taken
to be fixed, as in the original definition of the Ising
limit in Eq.~(\ref{e1.2}). However, if the restriction
that $\alpha$ be fixed is relaxed and $\alpha$ is allowed to
grow with $g$ in such a way that $\tilde\epsilon$ in Eq.~(\ref{e5.23})
remains small, then our results for the Green's functions in this
modified Ising limit remain valid in the larger range $2<D<4$.
Unfortunately, we still cannot extend the range of these results to the
physically important case $D=4$.

Nevertheless, in the range of dimension $0\leq D<4$ with $D\neq 2$ we have the
following picture. The scalar theory in this Ising-like
regime is very simple. The only remnant of the theory is a
renormalized mass $M_{\rm R}$, which approaches infinity,
and a one-point Green's function, which is the expectation
value of the scalar field. The higher Green's functions are all
negligible in this regime. If these results could be extended to
$D=4$,  we would have the equivalent of the Higgs phenomenon
without requiring the existence of a (so-far unobserved)
finite-mass Higgs particle.

\section*{ACKNOWLEDGMENTS}
\label{s6}
CMB and HFJ thank the Rockefeller Foundation for their hospitality and
support at the Bellagio Study and Conference Center.
CMB is grateful to the Theoretical Physics Group at Imperial College, London,
for their hospitality and he thanks the Fulbright Foundation and the PPARC for
financial support. CMB, PNM, and STB thank the U.S.~Department of Energy for
financial support.

\appendix
\section{EXPANSION COEFFICIENTS FOR THE EFFECTIVE ACTION}
\label{sa}

In this appendix we summarize the formalism needed to construct the coefficients
$\Gamma_n$ in the expansion for the effective action. We begin with the formula
for the connected vacuum generating functional ${\cal W}[J]$ in the presence of
an external source $J$:
\begin{eqnarray}
{\cal W}[J]=\ln{\cal Z}[J]=\sum_{n=1}^\infty {1\over n!}\int\ldots\int dx_1
\ldots dx_n\, J(x_1)\ldots J(x_n) G_n(x_1,\ldots,x_n).
\label{ex.aa}
\end{eqnarray}

Then the effective action $\Gamma[\phi]$ is defined by
\begin{eqnarray}
\Gamma[\phi]={\cal W}[J]-\int dx'\,[\phi(x')-G_1]J(x'),
\label{ex.bb}
\end{eqnarray}
where the classical field $\phi(x)$ is the expectation value of the field
in the presence of the source $J(x)$:
\begin{eqnarray}
\phi(x)={\cal G}_1(x)={\delta{\cal W}[J]\over\delta J(x)}.
\label{ex.cc}
\end{eqnarray}
We use the notation ${\cal G}_n$ to represent the $n$-point Green's functions in
the presence of the source $J$. Note that
\begin{eqnarray}
G_n={\cal G}_n\big|_{J=0}.
\label{ex.dd}
\end{eqnarray}

We assume the following functional Taylor series form for the effective
action:
\begin{eqnarray}
\Gamma[\phi] = \sum_{n=1}^\infty {1\over n!}\int\cdots\int dx_1\cdots dx_n\,
[\phi(x_1)-G_1]\cdots [\phi(x_n)-G_1] \Gamma_n(x_1,\ldots,x_n).
\label{ex.ee}
\end{eqnarray}

To find the coefficients $\Gamma_n$ we must differentiate $\Gamma[\phi]$
repeatedly with respect to $\phi$. This process requires the chain rule
\begin{eqnarray}
{\delta\over\delta\phi(z)} = {\delta J(z')\over\delta \phi(z)}
{\delta\over\delta J(z')} = [{\cal G}_2(z,z')]^{-1}{\delta\over\delta J(z')}
\label{ex.ff}
\end{eqnarray}
and the identity
\begin{eqnarray}
{\delta\over\delta J(y)}[{\cal G}_2(x,z)]^{-1}=[{\cal G}_2(x,x')]^{-1}
{\cal G}_3(x',y,z')[{\cal G}_2(z',z)]^{-1},
\label{ex.gg}
\end{eqnarray}
where repeated arguments indicate integration. We obtain $\Gamma_1$ by
calculating
\begin{eqnarray}
{\delta\Gamma[\phi]\over\delta\phi(z)} &=& [{\cal G}_2(z,z')]^{-1}{\cal G}_1
(z')-J(z)-[\phi(z')-G_1] [{\cal G}_2(z,z')]^{-1}\nonumber\\
&=&-J(z)+G_1 \int dz'\,[{\cal G}_2(z,z')]^{-1},
\label{ex.hh}
\end{eqnarray}
where we have used $\phi(z') = {\cal G}_1(z')$. Hence, evaluating this result at
$J\equiv 0$, which corresponds to $\phi\equiv G_1$, we obtain the first
coefficient in the effective action:
\begin{eqnarray}
\Gamma_1(z)={\delta\Gamma[\phi]\over\delta \phi(z)}\Biggm|_{\phi\equiv G_1}
=G_1\int dz'\,[G_2(z,z')]^{-1}.
\label{ex.ii}
\end{eqnarray}
There is a simple graphical interpretation for this result. The first
coefficient in the effective action is merely the one-point Green's function
with its external leg amputated.

Next, we calculate the second term in the effective action:
\begin{eqnarray}
{\delta^2\Gamma[\phi]\over\delta\phi(z)\delta\phi(u)} &=&
-[{\cal G}_2(z,u)]^{-1} + G_1 [{\cal G}_2(u,u')]^{-1}
\int dz'\,{\delta\over\delta J(u')}[{\cal G}_2(z,z')]^{-1}\nonumber\\
&=&-[{\cal G}_2(z,u)]^{-1} - G_1 \int dz'\, [{\cal G}_2(u,u')]^{-1}
[{\cal G}_2(z,a')]^{-1} [{\cal G}_2(z',b')]^{-1} {\cal G}_3(a',b',u').
\label{ex.jj}
\end{eqnarray}
Hence,
\begin{eqnarray}
\Gamma_2(u,z)=-[G_2(u,z)]^{-1}-[G_2(z,a')]^{-1}[G_2(u,u')]^{-1}
G_3(a',b',u') G_1\int dz'\, [G_2(z',b')]^{-1}.
\label{ex.kk}
\end{eqnarray}
The interpretation of this result is straightforward. The first term is the
usual one obtained in a parity symmetric theory. The second term is the
three-point Green's function with two of its legs amputated and its third
leg joined via a two-point Green's function to a one-point Green's function.

By continuing to differentiate, we can obtain the higher coefficients in the
effective action. Using an abbreviated notation we give the third coefficient
of the effective action:
\begin{eqnarray}
\Gamma_3=G_3(G_2)^{-3}- G_1G_4(G_2)^{-4}+3G_1(G_3)^2(G_2)^{-5},
\label{ex.ll}
\end{eqnarray}
where the numerical coefficient 3 indicates that there are three ways to
label the legs of the Green's functions. The fourth term is
\begin{eqnarray}
\Gamma_4 &=& G_4(G_2)^{-4}-3(G_3)^2(G_2)^{-5}-G_1G_5(G_2)^{-5}\nonumber\\
&&\qquad +10G_1G_3G_4 (G_2)^{-6}-15G_1 (G_3)^3(G_2)^{-7}.
\label{ex.mm}
\end{eqnarray}
Note that, in general, the terms not containing $G_1$ are the usual effective
action for a symmetric theory and that the terms containing $G_1$ are new
terms reflecting the broken parity symmetry.

The next two coefficients in the effective action are
\begin{eqnarray}
\Gamma_5 &=& G_5(G_2)^{-5}-10G_3G_4(G_2)^{-6}+15(G_3)^3(G_2)^{-7}
-G_1[G_6(G_2)^{-6} \nonumber\\
&&\quad -15G_3G_5 (G_2)^{-7}-10 (G_4)^2(G_2)^{-7}+105(G_3)^2 G_4(G_2)^{-8}
-105(G_3)^4(G_2)^{-9}]
\label{ex.nn}
\end{eqnarray}
and
\begin{eqnarray}
\Gamma_6 &=& G_6(G_2)^{-6}-15G_3G_5(G_2)^{-7}-10 (G_4)^2(G_2)^{-7}
+105 G_4(G_3)^2(G_2)^{-8} -105(G_3)^4(G_2)^{-9}\nonumber\\
&&\quad -G_1\left[G_7(G_2)^{-7}-21G_3G_6 (G_2)^{-8}-35 G_4 G_5(G_2)^{-8}
+210 (G_3)^2G_5 (G_2)^{-9}\right.\nonumber\\
&&\quad \left.+280G_3(G_4)^2(G_2)^{-9}
-1260G_4(G_3)^3(G_2)^{-10}+945(G_3)^5(G_2)^{-11}\right],
\label{ex.oo}
\end{eqnarray}
which have simple representations in terms of tree-graphs.

\end{document}